\documentclass[twocolumn,english,aps,prl,groupedaddress,superscriptaddress,longbibliography]{revtex4-2}
\usepackage{graphicx,epsfig,units}
\usepackage{xcolor} 
\usepackage{soul} 
\usepackage{amsmath,amsfonts,mathrsfs,amsbsy,bm,babel}
\usepackage[colorlinks=true,linkcolor=blue,citecolor=blue]{hyperref}

\begin{document}

\title{Role of Magnetic Defects and Defect-engineering of Magnetic Topological Insulators}
\author{Farhan Islam}
\affiliation{Ames National Laboratory, Ames, IA 50011, USA}
\affiliation{Department of Physics and Astronomy, Iowa State University, Ames, IA 50011, USA}

\author{Yongbin Lee}
\affiliation{Ames National Laboratory, Ames, IA 50011, USA}

\author{Daniel~M.~Pajerowski}
\affiliation{Neutron Scattering Division, Oak Ridge National Laboratory, Oak Ridge, TN 37831, USA}
\author{Wei Tian}
\affiliation{Neutron Scattering Division, Oak Ridge National Laboratory, Oak Ridge, TN 37831, USA}
\author{Jiaqiang Yan}
\affiliation{Materials Science and Technology Division, Oak Ridge National Laboratory, Oak Ridge, TN 37831, USA}

\author{Liqin Ke}
\affiliation{Ames National Laboratory, Ames, IA 50011, USA}

\author{Robert~J.~McQueeney}
\affiliation{Ames National Laboratory, Ames, IA 50011, USA}
\affiliation{Department of Physics and Astronomy, Iowa State University, Ames, IA 50011, USA}

\author{David~Vaknin}
\affiliation{Ames National Laboratory, Ames, IA 50011, USA}
\affiliation{Department of Physics and Astronomy, Iowa State University, Ames, IA 50011, USA}

\begin{abstract}
Magnetic defects play an important, but poorly understood, role in magnetic topological insulators (TIs). For example, topological surface transport and bulk magnetic properties are controlled by magnetic defects in Bi$_2$Se$_3$-based dilute ferromagnetic (FM) TIs and MnBi$_2$Te$_4$ (MBT)-based antiferromagnetic (AFM) TIs.  Despite its nascent ferromagnetism, our inelastic neutron scattering data show that a fraction of the Mn defects in Sb$_2$Te$_3$ form strong AFM dimer singlets within a quintuple block. The AFM superexchange coupling occurs via Mn-Te-Mn linear bonds and is identical to the AFM coupling between antisite defects and the FM Mn layer in MBT, establishing common interactions in the two materials classes. We also find that the FM correlations in (Sb$_{1-x}$Mn$_x$)$_2$Te$_3$ are likely driven by magnetic defects in adjacent quintuple blocks across the van der Waals gap. In addition to providing answers to long-standing questions about the evolution of FM order in dilute TI, these results also show that the evolution of global magnetic order from AFM to FM in Sb-substituted MBT is controlled by defect engineering of the intrablock and interblock coupling.
 \end{abstract}
\maketitle

\section{Introduction}

Spanning from the Kondo effect, to superconducting pair-breaking \cite{Lee2006}, to dilute magnetic semiconductors, and even entangled qubits \cite{Kitaev2003,Kitaev2006a}, magnetic defects and impurities can play both beneficial and deleterious roles in quantum materials. Thus, attempts to understand and control the influence of magnetic defects is key to unlock new quantum states of matter.  Perhaps nowhere is this more important than in topological insulators (TIs) such as Bi$_2Ch_3$ ($Ch=$ Se,Te), where collective ferromagnetic ordering of magnetic defects has been shown to break time-reversal symmetry, resulting in  dissipationless edge-currents through quantum anomalous Hall (QAH) effect \cite{Chang2013,Chen2010,Hsieh2009}. Nonetheless, the very nature of magnetism in these dilute and disordered magnetic TIs remain elusive and present a roadblock to their development as  platforms for topological devices.

\begin{figure}[b]
\includegraphics[width=0.6\linewidth]{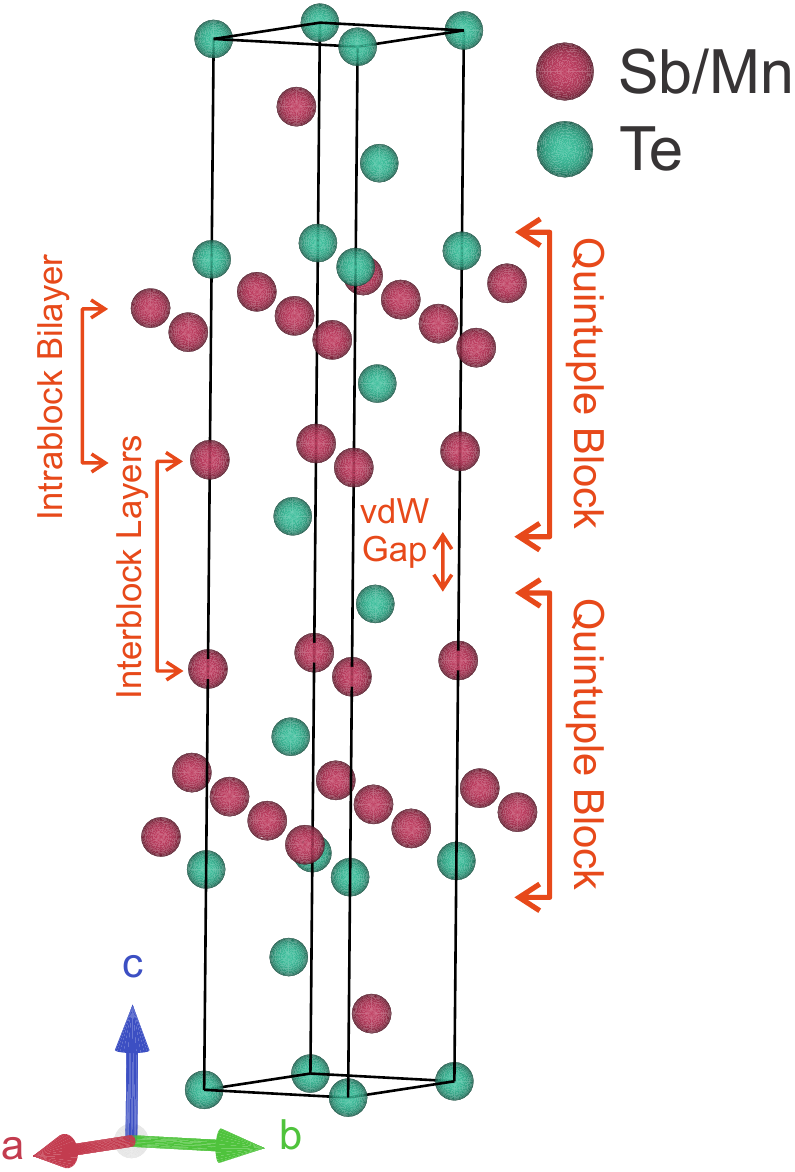}
\caption{\footnotesize {\bf Structure of Sb$_2$Te$_3$:} The unit cell of Sb$_2$Te$_3$ ($R\bar{3}m$ space group \#166) consists of three quintuple blocks, each separated by a van der Waals (vdW) gap. These blocks consist of Te-Sb-Te-Sb-Te layers. In the manuscript, we refer to the two Sb/Mn layers within the quintuple block as an {\it intrablock} bilayer. We also use the term {\it interblock} to refer to the adjacent Sb/Mn layers in different quintuple blocks across the vdW gap.}
\label{excitationPeaks}
\end{figure}

Recent realization of the magnetic TI MnBi$_2$Te$_4$(Bi$_2$Te$_3$)$_n$ (MBT), where intrinsic ferromagnetic (FM) Mn-layers are intercalated into Bi$_2$Te$_3$, provides a new route to QAH and axion insulators that depend on the FM or antiferromagnetic (AFM) nature of the interlayer coupling \cite{Yan2019,Otrokov2019,He2020,Li2019,Hu2020,Gong2019,Wu2020,Deng2021,Liu2020,Zhao2021,Ovchinnikov2021,Zhang2019}.
In addition, MBT and its sister compound MnSb$_2$Te$_4$ (MST) are known to contain various types of defects \cite{Yan2022}. Antisite Mn defects that occupy Bi or Sb sites (creating magnetic vacancies in the Mn layer) introduce new AFM couplings within a septuple block (intrablock) that lead to defect-driven ferrimagnetism \cite{Liu2021,Lai2021}. The exchange coupling between antisite defects and surface electronic bands has been shown to be a pivotal factor that controls the surface magnetism and consequently the Chern gap \cite{Liu2021,Garnica2022,Shikin2021,Sitnicka2022,Liu}. These studies reveal the important role of Mn defects in MBT and MST, and the prospect that defect-engineering can control and tailor the global magnetic ground state that affect topological surface states.
This accentuates the importance of unraveling the physics of dilute magnetic TIs, such as (Bi$_{1-x}$Mn$_{x}$)$_2$Te$_3$ or (Sb$_{1-x}$Mn$_{x}$)$_2$Te$_3$, where, in a similar fashion, Mn randomly occupies the Bi/Sb sites \cite{Watson2013,Hor2010,Henk2012,Choi2004,Ghasemi2016,Vaknin2019}.  Questions regarding the origin of the magnetic couplings, as  localized superexchange or long-range exchange via carriers (i.e. RKKY) are still open.

Here, we approach these questions from the dilute end of the spectrum by employing magnetic inelastic neutron scattering (INS) measurements on single crystals of  (Sb$_{0.97}$Mn$_{0.03}$)$_{2}$Te$_{3}$. This dilute TI is on the verge of FM order, despite our observation of the dominant AFM intrablock coupling between Mn in different Sb layers. In particular, we find Mn-Mn singlets are formed from strong AFM superexchange interactions mediated through Mn-Te-Mn linear bonds. This configuration is nearly identical to the AFM coupling between antisite Mn defects and the Mn layers that give rise to ferrimagnetism in MBT and MST. We also find evidence at very low energies ($\approx 0.1$ meV) for nascent FM correlations that drive long-range FM order in dilute TIs.  Surprisingly, INS data indicate that this FM coupling occurs between rafts of Mn ions on either side of the van der Waals (vdW) gap, providing a glimpse of how FM long-range order evolves from the dilute limit. This result highlights the role that defects play in mediating the interblock interactions that control the crossover from AFM order in MBT to FM order in MST, for which antisite defect-concentrations are higher\cite{Liu2021,Riberolles2021}.

\begin{figure}
\includegraphics[width=1\linewidth]{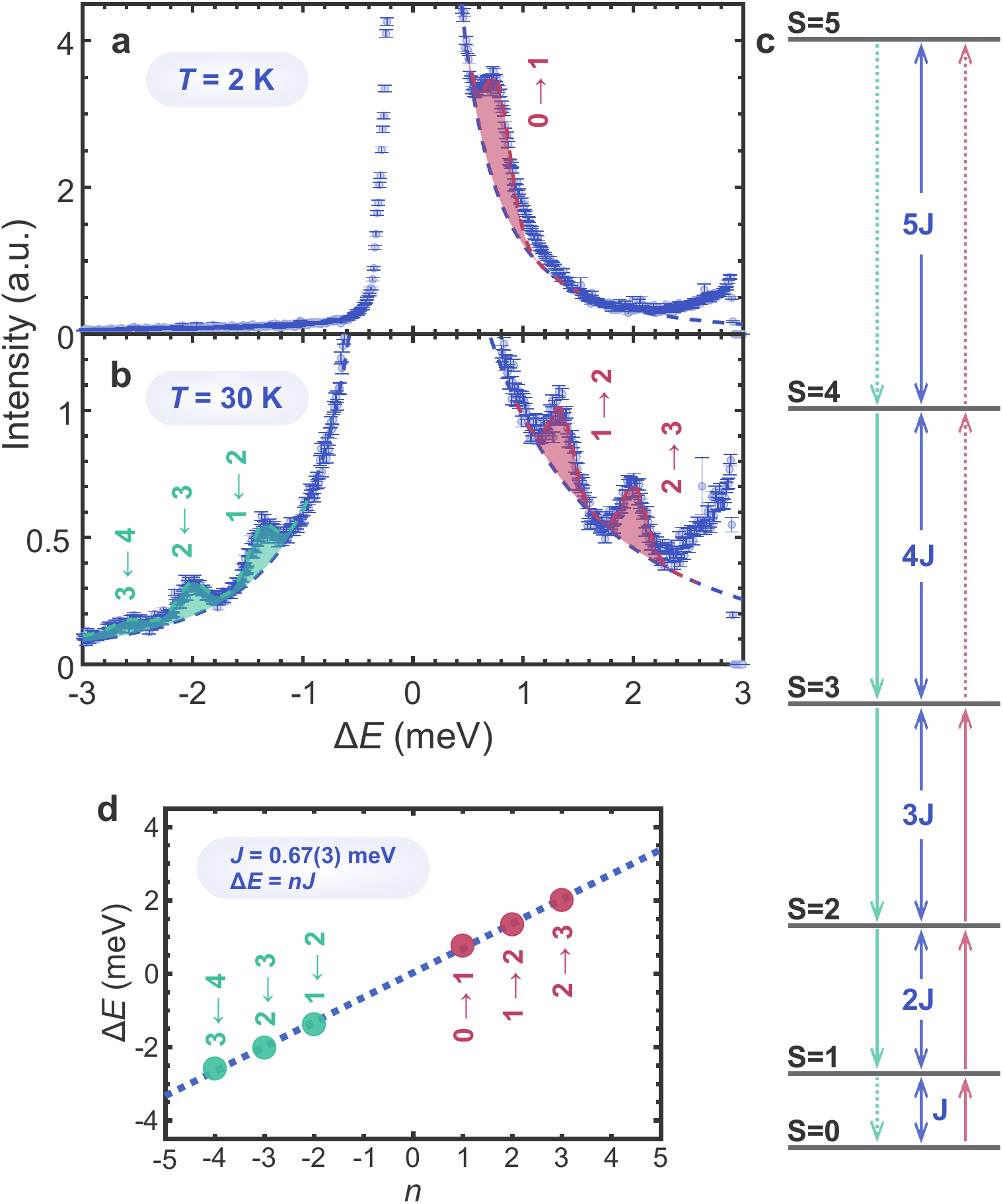}
\caption{\footnotesize {\bf Spectroscopic evidence of an AFM Mn-Mn dimer:} Integrated intensity of the AFM dimer spectrum obtained by summing up two symmetrically equivalent regions of interest (ROI): one with $H = [-0.25, 0.25]$ and $L = [2.5, 6]$; the other with $H = [0.2, 0.9]$ and $L = [-4, 2]$ at {\bf a} $T=2$ K and {\bf b} $T=30$ K.  Fitted spin-state transition peaks are shaded and labeled by their initial and final spin-states. {\bf c,} The level spectrum of a $s = \frac{5}{2}$ AFM dimer with arrows showing the energies of dipole-allowed transitions with $\Delta S=\pm1$ observable by INS (excitations with dotted lines are not observed in our experiment). {\bf d,} Transition energies as a function of $n$ (see text) with a linear fit that determines the AFM coupling stength $J=0.67(3)$ meV.}
\label{excitationPeaks}
\end{figure}

\begin{figure*}
\includegraphics[width=0.83\linewidth]{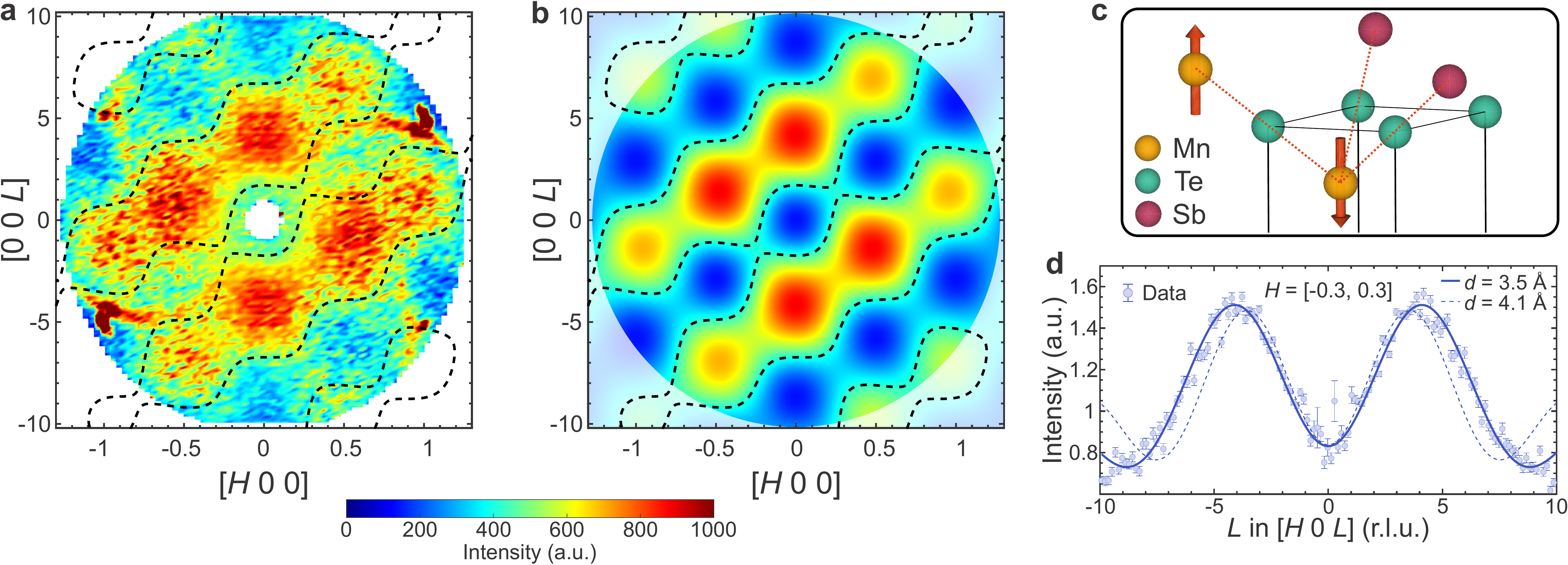}
\caption{\footnotesize {\bf Configuration of the AFM Mn-Mn intrablock dimer:} {\bf a,} AFM dimer structure factor for (Sb$_{0.97}$Mn$_{0.03}$)$_{2}$Te$_{3}$ measured at $T$  = 2 K with incident neutron energy $E_i$ = 3.32 meV. Data in the $(H,0,L)$ plane is obtained by averaging over $K = [-0.1, 0.1]$ r.l.u. range and an energy transfer range $\Delta E = [0.6, 0.9]$ meV, corresponding to the $0\rightarrow1$ spin-state transition. {\bf b,} Structure factor calculations obtained by averaging the three symmetry-equivalent intrablock bilayer AFM dimers residing at NNN positions, as shown in {\bf c}. Calculations are performed by spherically averaging over anti-parallel moments. {\bf c,}  Red spheres represent Sb sites that are substituted randomly by Mn (yellow spheres).  The intrablock bilayer region shows the three equivalent NNN dimers coupled via Te atom at a Mn-Te-Mn 180$^\circ$ bond-angle (red dashed line). Moment directions are merely for representation. {\bf d,} Line-cut along the $L$ direction integrated over $H = [-0.3, 0.3]$ on the slice shown in {\bf a}. The solid line is the best fit obtained by varying the bilayer spacing ($d=3.5$ \AA). The dashed line is a similar fit with a fixed bilayer spacing of $d=4.1$ \AA~as determined from the reported crystal structure of Sb$_2$Te$_3$. This indicates a local contraction of the Mn-Te-Mn bond-length compared to the unsubstantiated Sb-Te-Sb.}
\label{sf}
\end{figure*}

\section{Dimer Formation}

Figure\ \ref{excitationPeaks}(a) and (b) show INS spectra at $T=2$ K and $T=30$ K of (Sb$_{0.97}$Mn$_{0.03}$)$_{2}$Te$_{3}$.  At $T= 2$ K [Fig. \ref{excitationPeaks}(a)], only one excitation peak is observed in the neutron energy loss channel at $\Delta E \simeq 0.7$ meV.  As the temperature is raised to $T =30$ K [Fig.\ \ref{excitationPeaks}(b)], a series of excitations develops on both the energy-loss and the energy-gain channels. The sharp peaks are fitted to Gaussian lineshapes including background, showing that the peak positions are equally spaced.

Here, we show that these excitations comprise the spin-state transitions of an AFM Mn-Mn dimer. Assuming $\mbox{\bf {s}}_i$ to be the spin operator of a Mn$^{2+}$ atom  with $s=5/2$, the Hamiltonian for an isolated dimer is:
\begin{equation}
    H = J\mbox{\bf{s}}_1\cdot\mbox{\bf{s}}_2
\end{equation}
With AFM exchange coupling, $J>0$, the ground state of the coupled dimer is a singlet with total spin $S=0$ and energy levels given by: 
\begin{equation}
    E(S) =J/2\left[S(S+1)\right],
\label{peakPosition}
\end{equation}
where $S = \{0,1,2,3,4,5\}$ is the total spin of the dimer, as shown in Fig.~\ref{excitationPeaks}(c). INS measures the dimer state transitions with selection rule that $\Delta S = 0,\pm1$. Thus, inelastic excitations observed by INS probe transitions only between adjacent energy-levels, corresponding to a spectrum of equally spaced excitations with $\Delta E = \pm nJ$ and $n = 1,2,3,4,5$ (see more details in the  SM \cite{SM}). In Fig.~\ref{excitationPeaks}(d), the transition energies are fitted to a linear $\Delta E = nJ$ function, yielding an AFM $J = 0.67(3)$ meV. For $k_BT \ll J$, as in Fig.~\ref{excitationPeaks}(a), the single excitation at $\Delta E=J$ confirms an AFM exchange coupling with a singlet ground state. Analysis of the suppression of the bulk susceptibility at low temperatures also provides evidence for AFM singlet formation \cite{SM}. Our estimate of the portion of Mn ions participating in dimer formation is 12(2)\%, which is comparable to the statistically calculated distribution of Mn-Te-Mn configuration ($\approx 8$\%) (see SM \cite{SM}).

Establishing the existence of AFM dimers, we proceed to determine their bonding configuration by analyzing the $Q-$dependence of the neutron intensity. Figure \ref{sf}(a) shows the measured dimer structure factor of the $0
\rightarrow1$ transition in the ($H,0,L$) plane at $T = 2$ K, as obtained by integrating over $\Delta E$ = [0.6, 0.9] meV. To model the structure factor, we use the following:
\begin{equation}
    I_{FF} =C|f(Q)|^2\left|{\sum_{i=1}^{k}}{\rm e^{i{\bf Q\cdot r_i}}{\bf {\hat Q}}\times({\bf {\hat m_{\rm i}}}\times{\bf \hat{Q}}})\right|^2,
\label{sfequation}
\end{equation}
where the scattering vector ${\bf Q}=H{\bf a^*}+K{\bf b^*}+L{\bf c^*}$, $\bf{\hat Q}$ is its unit vector, $\bf{\hat{m}}$ is the unit vector of the magnetic moment, and $f(Q)$ is the Mn$^{2+}$ form factor\cite{Brown2006}. Using Eq. (\ref{sfequation}) with $k=2$, we tested various dimer bonds that correspond to Mn substitutions on the Sb sublattice and compared them to Fig.~\ref{sf}(a).  Our best fit to the data is achieved for dimers where intrablock Mn-Mn bonds form between NNN sites in adjacent Sb layers, as depicted in Fig.~\ref{sf}(c). The oblique orientation of intensity maxima in Figs.~\ref{sf}(a) and (b) is consistent with the NNN bond vector having a component along both c-direction and ab-plane.

Closer examination of the structure surrounding NNN Mn-Mn dimers in Fig.~\ref{sf}(c) reveals that they are coupled through a Te atom with Mn-Te-Mn 180$^\circ$ bond-angle. It has been established that magnetic coupling between transition metal ions via chalcogen $p$-orbitals (O, S, Se, Te) in a linear bond configuration yields an AFM superexchange interaction, following the Anderson-Goodenough-Kanamori rules \cite{Anderson1963-b,Goodenough1963,Kanamori1959}. We note that this result is consistent with the observation of similar linearly-bonded Mn-Te-Mn AFM dimers in the cubic Sn$_{0.95}$Mn$_{0.05}$Te magnetic TI \cite{Vaknin2020}. 

The intensity maxima of line-cuts along $L$ with $H$ in the $[-0.3, 0.3]$ range, as shown in Fig.~\ref{sf}(d), are sensitive to the apparent spacing between Sb-Sb layers that host the AFM dimer. Our detailed fits show that the bilayer spacing, as measured by the Mn-Mn dimer structure factor, is contracted from 4.1 {\AA} of the stoichiometric bilayer spacing to $d = 3.50(3)$ \AA) in this dilute Mn-doped system. Assuming no structural change in the $ab$-plane, the effective Mn-Te-Mn bond-length is 6.04(3) \AA. We confirm, using first-principles calculations described below, that this is caused by local structural relaxation of the Mn-Mn dimer adopting a similar apparent bilayer spacing as found in the hexagonal MnTe ($d\approx$ 3.4 \AA) \cite{Szuszkiewicz2006}.

\section{Ferromagnetic Correlations}

\begin{figure*}
\includegraphics[width=1\linewidth]{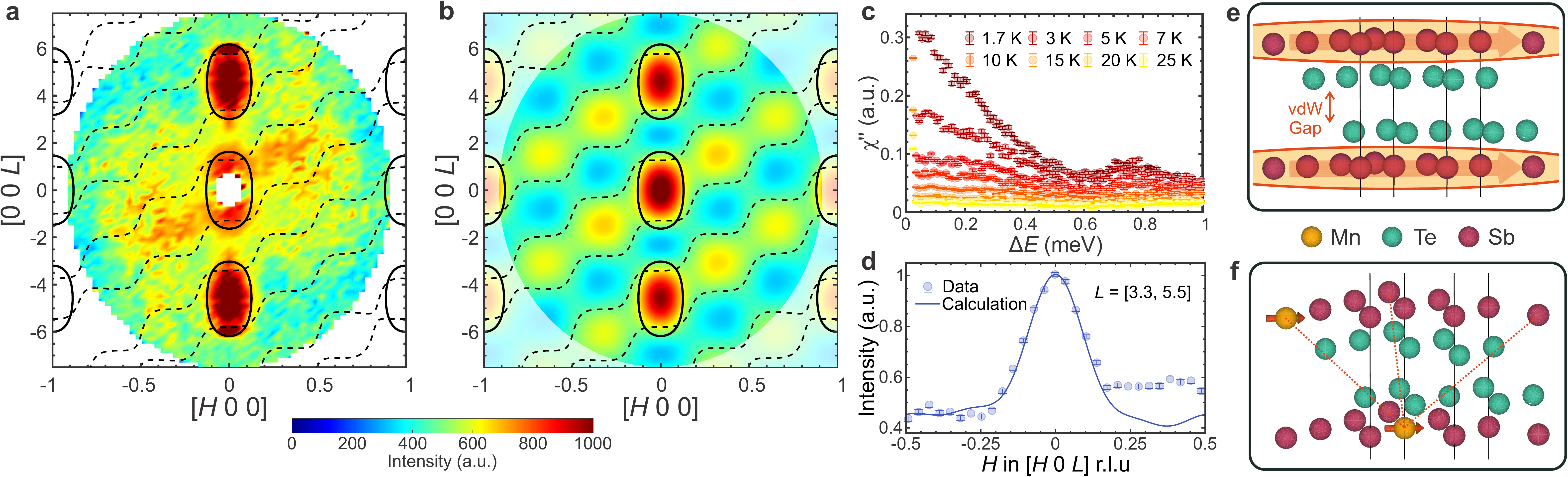}
\caption{\footnotesize {\bf Ferromagnetism across the van der Waals gap:} {\bf a,} The structure factor of the developing FM correlations in (Sb$_{0.97}$Mn$_{0.03}$)$_{2}$Te$_{3}$ as measured at $T$  = 2 K and an incident neutron energy $E_i$ = 1.55 meV. Data shown is integrated over $\Delta E = [0.04, 0.1]$ meV and $K = [-0.058, 0.058]$ r.l.u. {\bf b,} Structure factor calculations obtained by summing up the contributions from FM interlayer correlations (solid contour lines) and FM dimers (dashed contour lines), both coupled across the vdW gap. Calculations are performed by spherically averaging over parallel moments. {\bf c,}  The decreasing dynamical susceptibility ($\chi^{\prime\prime}$) of the spectra of FM correlations with increasing temperature, obtained by extracting the energy spectra by summing over $H$ = [-0.1, 0.1] r.l.u. and $L$ = [3.3, 6.3] r.l.u., implies critical fluctuations of a nearly FM system. {\bf d,} $H$-cut obtained by summing up over $L$ = [3.3, 5.5] in the data ({\bf a}) and calculation ({\bf b}) shows that the width is best fit by assuming a correlation volume of $k=5\times5\times2=50$ sites. {\bf e,} Depiction of interblock FM correlations across the vdW gap consisting of two FM layers. {\bf f,} Illustration of the FM dimers formed across the vdW gap. Closer examination of the dimers show that Mn-Mn are coupled through a nearly linear Mn-Te-Te-Mn bond. Moment directions are merely for representation.}
\label{sfcluster}
\end{figure*}

The INS data also manifest quasielastic fluctuations at lower energies $\Delta E \approx 0.1$ meV that indicate the development of long-range FM order, as shown in Fig. \ref{sfcluster}. The decreasing dynamical susceptibility, $\chi''(E)=I(1-{\rm exp}(-E/kT))$, with increasing temperature suggests critical fluctuations of a nearly FM system. Below 0.1 meV, Fig.~\ref{sfcluster}(a) shows that the quasielastic response is strongly peaked at $\bf{Q}$ $\approx (0,0,5)$ with a relatively narrow profile in the $H$-direction.  The centering of FM fluctuations near $L=5$ is evidence that this feature arises from FM interblock correlations between Sb layers \textit{across} the vdW gap. Fits to the $H$-cut in Fig. \ref{sfcluster}(d) indicate an intralayer FM correlation length of $\approx$ 5 in-plane unit cells. 

To quantitatively determine the origin of quasielastic FM fluctuations in Fig. \ref{sfcluster}, we simulate the scattering using Eq.~(\ref{sfequation}) and assume that the major contribution to the scattering is from magnetization densities coupled across the vdW gap, as shown in Fig. \ref{sfcluster}(e). The $Q$-dependence of the scattering in Fig. \ref{sfcluster}(a) can be obtained from a model of FM dimer correlations across the vdW gap.  Similar to the AFM dimer, the FM dimer contributes to the weak diffused signal seen along the diagonal in Fig.~\ref{sfcluster}(a), indicating that it originates from dimer bond vectors having a particular angle with respect to the basal planes. We propose that the FM interblock dimers are formed by pairs of Mn atoms that are connected through a nearly linear Mn-Te-Te-Mn bond across the vdW gap, as shown in Fig. \ref{sfcluster}(f). Figure \ref{sfcluster}(b) shows calculated structure factor of good qualitative agreement with the experimental observation in Fig.~\ref{sfcluster}(a). The solid contour lines in Fig.~\ref{sfcluster}(a) and (b) represent the signal from FM correlations within a single Sb layer and the dashed lines represent the signal from the FM dimers.
From inelastic data shown in Fig. \ref{sfcluster}, it is not possible to ascertain the nature of the intralayer FM correlations. The 3\% Mn concentration of our sample is below the threshold for long-range FM order, and we would expect only approximately two Mn ions within the correlation volume of $5\times 5 \times 2 = 50$ sites on average.  This suggests that the development of FM long-range order requires a combination of Mn clustering and/or long-range intralayer couplings.


\begin{figure}
\includegraphics[width=0.9\linewidth]{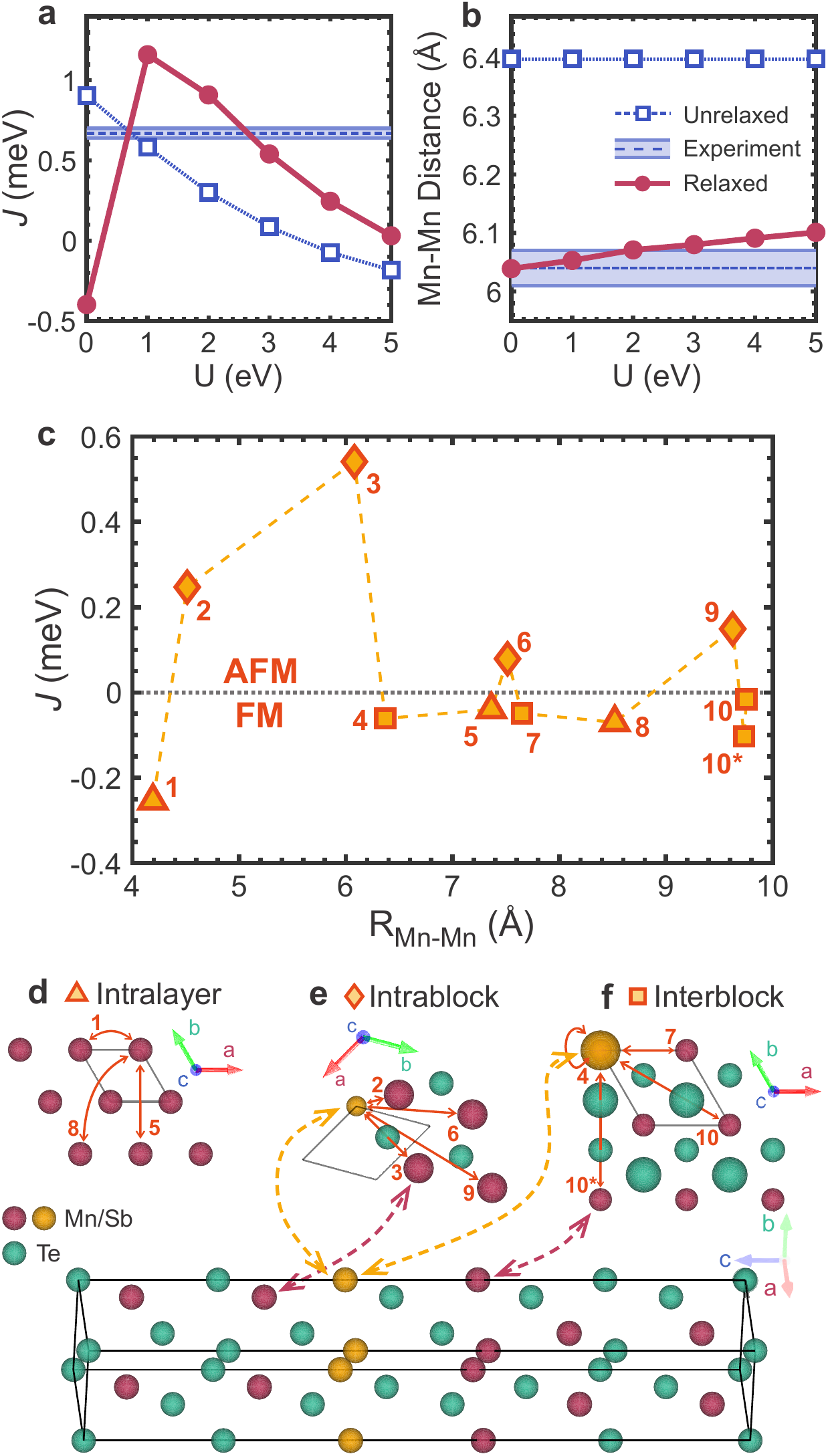}
\caption{\footnotesize {\bf Density functional theory calculation of various Mn-Mn pairwise interactions in a supercell:} {\bf a,} Exchange coupling $J$ of the dominant AFM intrablock Mn-Te-Mn dimer as a function of $U$ and compared with experiments. The solid red circles and hollow blue squares represent the calculated exchange coupling of relaxed and unrelaxed states, respectively. The shaded blue region is the experimentally determined exchange coupling of the Mn-Te-Mn dimer. {\bf b,} Calculated Mn-Te-Mn (of the AFM dimer) length as a function of $U$. {\bf c,} Exchange coupling constants $J$ as a function of various Mn-Mn pair-length in the supercell calculated at $U=3$ eV. {\bf d--f,} Illustration of the various Mn-Mn pairs, labeled corresponding to the ones in {\bf c}. Mn/Sb atoms in different colors refer to atoms in different layers within the block.}
\label{DFTfigure}
\end{figure}

\section{Theoretical Considerations}

We use first-principles calculations of the magnetic interactions to confirm our main discoveries of strong intrablock AFM dimer formation and the key intralayer and interblock FM interactions that drive FM order. In doing so, we recognize that electron correlations can substantially affect the magnetic interactions in vdW materials~\cite{Ke2021}.
DFT+$U$ is a simple form to account for on-site non-local correlation effects.
However, the choice of $U$ value can be ambiguous, and therefore we resort to the experimental result to determine it.
Sizable $U$ values of 4 -- 5 eV in DFT+$U$ calculations yield good descriptions of magnetic interactions in MnBi$_{2}$Te$_{4}$ and MnSb$_{2}$Te$_{4}$ systems~\cite{Li2020,Li2021a,Riberolles2021,Lai2021}.
To determine $U$ in this dilute system, we compare the experimentally determined $J$ of the strongest AFM Mn-Te-Mn intrablock dimer interaction to the calculations.
Figure \ref{DFTfigure}(a) compares the $U$ dependence of calculated exchange constant $J$ to the experimentally determined value of 0.67 meV. A reasonable value of $U=3$ eV is chosen for subsequent calculations.

Figure \ref{DFTfigure}(b) shows the length of Mn-Te-Mn dimer as a function of $U$. We find that the relaxed length is shorter than the unrelaxed one for all $U$'s, consistent with the experimentally observed contracted bond length discussed above. We also note that, the Mn-Te-Mn dimer length is practically insensitive to the value of $U$.

Figure \ref{DFTfigure}(c) shows the calculated exchange interactions as a function of distance between various Mn-Mn pairs for $U=3$ eV.
Figure \ref{DFTfigure}(d--f) depicts and labels the various Mn-Mn dimers in the calculated supercell. The DFT calculation show that all in-plane and interblock nearest neighbors (NN) are ferromagnetically coupled, as observed experimentally for the FM cluster across the vdW gap. The strongest calculated FM interblock interaction is formed via a linearly connected Mn-Te-Te-Mn dimer. The DFT calculations also predict that all the intrablock interactions are AFM, with  the strongest one being the interaction formed via the same Mn-Te-Mn bond discussed above. In the SM \cite{SM}, we show that these observations are consistent with the DFT calculations with $U=2, 3,4$ eV.

\section{General Implications to MST/MBT and Beyond}

Although a dilute system, (Sb$_{0.97}$Mn$_{0.03}$)$_{2}$Te$_3$ is an important material with broader implication to magnetic TIs and beyond. In both MBT and MST, it is now clear that antisite mixing between Mn and Bi/Sb occurs at levels ranging from a few percent up to $\sim$~15\% \cite{Murakami2019,Riberolles2021,Yan2019,Liu2021,Lai2021,Du2021}, affecting the magnetic ground state of these systems. The Bi/Sb layers host low concentrations of Mn spin-defects with interactions that can be understood from our studies of Mn in dilute parent compounds. Indeed, it has been inferred that antisite Mn spins are coupled antiferromagnetically to the main Mn layer, resulting in defect-driven ferrimagnetism. Furthermore, high-field magnetization data estimate this AFM coupling to be 0.8 meV in MST which is consistent with the value found here (0.67(3) meV) for the AFM Mn-Mn dimers. We now know that this AFM interaction occurs in both dilute and intercalated TIs through superexchange via linear Mn-Te-Mn bonds within the septulple or quintuple blocks.

\begin{figure}
\includegraphics[width=0.8\linewidth]{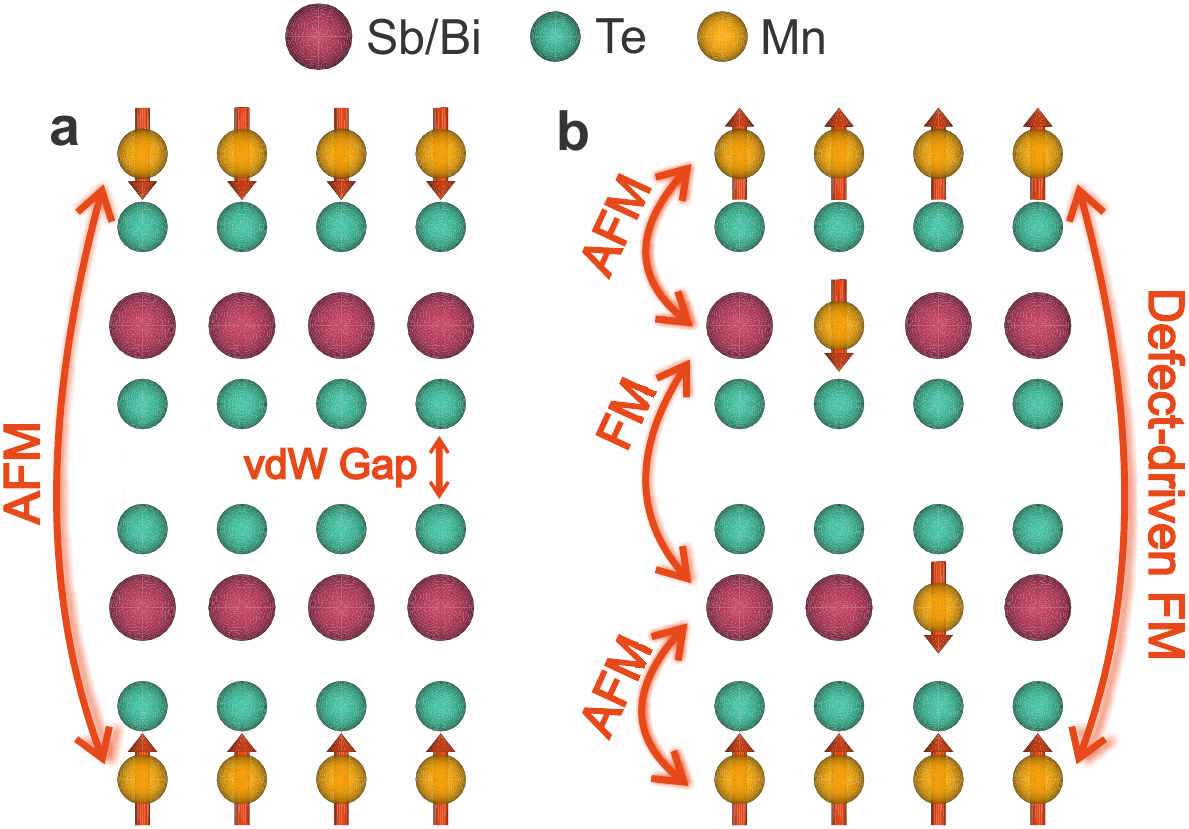}
\caption{\footnotesize {\bf AFM to FM path mediated via defects in MBT and MST:} {\bf a,} Projection through an in-plane direction of MST/MBT system showing the AFM magnetic structure. {\bf b,} The role of antisite Mn defects in the Sb sites in transforming the AFM to FM ground state. The mechanism is depicted by orange arrows. The Mn-defect on the first Sb-layer from the bottom is coupled antiferromagnetically with respect to the Mn-layer at the bottom due to superexchange. The Mn-defect on the following Sb-layer above is ferromagnetically coupled with respect to the one below, consistent with the DFT results and experimentally observed FM coupling across the vdW gap. Finally, the Mn-layers on the top are flipped to couple antiferromagnetically with respect to the Mn-defect on the Sb-layer below due to superexchange. Therefore, the Mn-defects are creating a pathway to manifest a ferromagnetic coupling for the Mn-layers in the MBT/MST system.}
\label{mbtConnection}
\end{figure}

Our observation of predominant FM interactions that occur across the vdW gap can also explain why either global FM or AFM order can be achieved in MST, which is found to be dependent on growth conditions that affect the defect configuration. At low antisite defect concentrations, as occurs in MBT, intrinsic interblock AFM coupling is dominant. Based on the results shown here, an increased concentration of antisite defects, as found in MST, promotes competing FM interblock coupling. Figure \ref{mbtConnection} illustrates likely scenario of how defects of the sort found in the dilute system can drive the ground state of MST from AFM to FM.

For dilute FM TIs, these results resolve some questions about the key interactions that drive long-range FM order, but also raise new  questions.  The formation of strong AFM singlets with a coupling strength exceeding all other magnetic interactions in the system seems to be in direct conflict with the development of FM order.  We find that a variety of materials, such as the aforementioned MBT and MnTe, are all characterized by similarly strong AFM coupling in linear Mn-Te-Mn bonding geometries. It is interesting to note that the AFM dimer singlets formed from linear Mn-Te-Mn bonds are also reported in another dilute FM TI, i.e., the face-centered cubic Sn$_{1-x}$Mn$_{x}$Te \cite{Vaknin2020}.  We can only assume that these singlets are magnetically dead and do not contribute to the long-range ferromagnetism.  Long-range order is then left to develop through Mn ions that do not participate in such dimers.  The essential FM interactions point to two conclusions.  First, they show evidence for long-ranged intralayer FM interactions with a correlation length of several unit cells, similar to the long-range interactions found in MBT \cite{Li2020}. This suggests that intralayer interactions may be mediated by conduction electrons, although we can not rule out a contribution from intralayer NN FM superexchange coupling via Mn-Te-Mn with bond angles that are closer to 90$^\circ$. Second, evidence for FM coupling across the vdW gap has a contribution from longer linear Mn-Te-Te-Mn chains.

Increased Mn concentration in dilute alloys does result in long-range order. Similar powder INS measurements on (Bi$_{0.95}$Mn$_{0.05}$)$_2$Te$_3$ have revealed a long-range FM ordering within the basal plane that exhibits collective two-dimensional (2D) magnetic excitations below $T_C \simeq$ 13 K \cite{Vaknin2019}. More generally, the INS methods applied to other dilute systems show great promise to unravel the variety of open questions in dilute magnetic systems, such as magnetic semiconductors, spin-glasses, and Kondo-effect materials.

\section{Methods}
\subsection{Crystal Growth and Characterization}
A large single-crystal of (Sb$_{0.97}$Mn$_{0.03}$)$_{2}$Te$_{3}$ weighing $\sim4$ grams was by Bridgeman method \cite{May2020}, crystallizing in the $R\bar{3}m$ space group (\#166), with lattice parameters $a = b = 4.26$ {\AA}, $c = 30.4$ {\AA} at $T=2$ K. Magnetic susceptibility measurements down to $T=3$ K do not show signatures of a magnetic ordering transition.  As shown in the Supplementary Material (SM) \cite{SM}, Curie-Weiss analysis of the susceptibility confirms the concentration of the Mn in the parent compound. 
 
\subsection{Inelastic Neutron Scattering}
The neutron scattering experiments were conducted using the Cold Neutron Chopper Spectrometer (CNCS) at the Spallation Neutron Source (SNS) at Oak Ridge National Laboratory\cite{Ehlers2016}. The single-crystal sample was attached to a rod that was dropped into liquid helium Dewar (an `orange' wet $^{4}$He cryostat). INS experiments were performed at the 2 -- 30 K temperature range. The data were collected with the `high-flux-mode' and fixed incident neutron energies of $E_{i}=$ 1.55, 3.32, and 12.00 meV, which have approximately gaussian full-width-half-maximum instrumental elastic resolutions of 0.04, 0.11, and 0.68 meV, respectively. Throughout the manuscript, data are described in terms of Miller indices $H$ and $L$ defined with respect to the hexagonal reciprocal lattice unit vectors (r.l.u.) {\bf a$^*$} and {\bf c$^*$}. The crystal was oriented with a horizontal ($H,0,L$) scattering plane and direction perpendicular to this plane is labeled by the hexagonal vector $(-K,2K,0)$. The MANTID software package \cite{Arnold2014} was used to reduce and concatenate time-of-flight data sets that were collected by incrementally rotating the crystal around the vertical ($-1,2,0$) axis.

\subsection{Density Functional Theorem}
Density functional theory (DFT) calculations are carried out to estimate the exchange interaction between magnetic impurities. Calculations are performed within the framework of the plane-wave projector-augmented wave formalism, as implemented in the Vienna ab initio simulation package~\cite{Kresse1996b,Kresse1999a}.
The Perdew-Burke-Ernzerhof generalized gradient approximation~\cite{Perdew1996a} is used for the exchange-correlation functional.
The Hubbard Coulomb interaction $U$ is included to better account for the strong correlation of Mn-$3d$ electrons~\cite{Dudarev1998a}.

To simulate a diluted magnetic impurity system, we construct 5$\times$4$\times$2 supercell of Sb$_2$Te$_3$ and substitute two Sb sites with Mn atoms.
The composition of the resulted Sb$_{78}$Mn$_{2}$Te$_{120}$ cell compares well with the experimental composition of (Sb$_{0.97}$Mn$_{0.03}$)$_{2}$Te$_{3}$.
In total, 11 configurations of Mn dimers are considered: three configurations with the two Mn dimer atoms situated in the same Sb layer (intralayer), four configurations with the two Mn atoms in the different Sb layer of a quintuple block (intrablock), and four configurations with the two Mn atoms in neighboring quintuple blocks across the vdW gap (interblock).
For each dimer configuration, the energies of FM and AFM ordering, $E_\text{FM}$ and $E_\text{AFM}$, respectively, are calculated to estimate the relative stability of dimer formation and the effective exchange coupling $J$.

The dilute Mn impurities induce local structural relaxation, which can affect magnetic interactions.
We optimize the local structures by relaxing the internal atomic coordinates until the force on each atom is less than 0.1 ${\rm meV/\AA}$ while keeping the lattice parameters and cell shape fixed, considering the small impurity concentration.
For the interblock dimer configurations, we also consider a density	functional that better describes the long-range electron correlations responsible for the vdW forces\cite{Grimme2006}.
The plane-wave cutoff energy was set to 350 eV, and the total energy was converged to $E-5$ eV with 2$\times$2$\times$2 $k$-mesh.

\section{Author contributions}
R.~J.~M.~and D.~V.~conceived and designed the experiments. J.-Q.~Y.~and W.~T.~synthesized the crystals and characterized them. F.~I.~, D.~M.~P.~, R.~J.~M., and D.~V.~conducted neutron scattering measurements. F.~I.~, R.~J.~M., and D.~V.~performed experimental data analysis and modeling. Y.~L.~and L.~K.~performed the DFT calculations. F.~I.~prepared the figures. F.~I., R.~J.~M., and D.~V.~wrote the manuscript. All coauthors read and commented on the final manuscript.

\section{Acknowledgments}
This research was supported by the U.S. Department of Energy, Office of Basic Energy Sciences, Division of Materials Sciences and Engineering. Ames Laboratory is operated for the U.S. Department of Energy by Iowa State University under Contract No. DE-AC02-07CH11358.  A portion of this research used resources at the Spallation Neutron Source, a DOE Office of Science User Facility operated by the Oak Ridge National Laboratory.
 This manuscript has been authored in part by UT-Battelle, LLC, under contract DE-AC05-00OR22725 with the US Department of Energy (DOE). The US government retains and the publisher, by accepting the article for publication, acknowledges that the US government retains a nonexclusive, paid-up, irrevocable, worldwide license to publish or reproduce the published form of this manuscript, or allow others to do so, for US government purposes. DOE will provide public access to these results of federally sponsored research in accordance with the DOE Public Access Plan (\url{http://energy.gov/downloads/doe-public-access-plan}).

\bibliography{Mag-TIs_zotero.bib}

\clearpage
\pagebreak

\setcounter{page}{1}
\setcounter{figure}{0}
\setcounter{equation}{0}
\setcounter{table}{0}
\setcounter{section}{0}

\renewcommand{\thefigure}{S\arabic{figure}}
\renewcommand{\theequation}{S\arabic{equation}}
\renewcommand{\thetable}{S\arabic{table}}
\renewcommand{\thepage}{S\arabic{page}}

{\center{ 
{\bf Supplemental Material \\
Role of Magnetic Defects and Defect-engineering of Magnetic Topological Insulators} \\~\\

{Farhan Islam,$^{1,2}$ Yongbin Lee,$^1$ Daniel~M.~Pajerowski,$^3$ Wei Tian,$^3$ Jiaqiang Yan,$^4$ Liqin Ke,$^{1}$ Robert~J.~McQueeney,$^{1,2}$ and David Vaknin$^{1,2}$}\\~\\
{\it $^1$ {Ames National Laboratory, Ames, IA 50011, USA} \\
$^2${Department of Physics and Astronomy, Iowa State University, Ames, IA 50011, USA} \\
$^3${Neutron Scattering Division, Oak Ridge National Laboratory, Oak Ridge, TN 37831, USA}\\
$^4${Materials Science and Technology Division, Oak Ridge National Laboratory, Oak Ridge, TN 37831, USA}\\
}
}
}

\section{Magnetic Susceptibility and Magnetization}

\begin{figure}[ht]
\includegraphics[width = 3.3in]{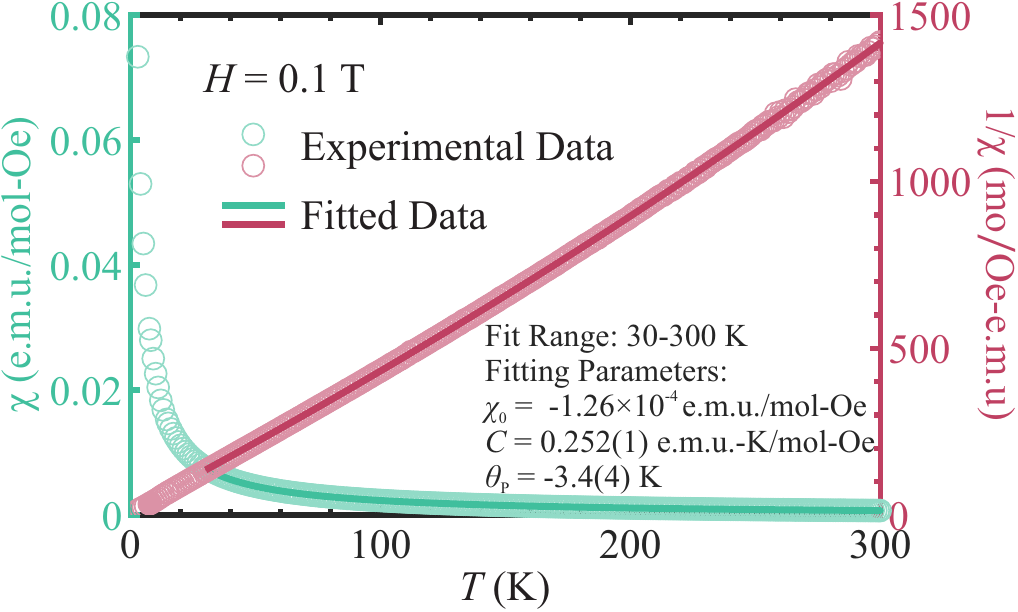}
\caption{{\bf Magnetic susceptibility:} Temperature dependence of the measured susceptibility and inverse susceptibility  for Sb$_{1.94}$Mn$_{0.06}$Te$_3$ along with a fit by Eq.~(\ref{CW}) in the 30 to 300 K temperature range in solid line.}
\label{MagSus}
\end{figure}

Figure\ \ref{MagSus} shows the measured temperature dependence of magnetic susceptibility ($\chi$) and its inverse ($1/\chi$). To model the susceptibility, we use three terms following the procedure outlined in the SM of Ref. \cite{Vaknin2020}. The three terms represent the diamagnetic contribution ($\chi_0$), the Curie-Weiss susceptibility ($\chi_{\rm cw}$), and the magnetic susceptibility expected from the dimers ($\chi_d$). At low temperatures, when a fraction ($f_d$) of the magnetic atoms form dimers, the total susceptibility is given by  
\begin{eqnarray}
\chi_{\rm {total}} & =& \chi_0 + (1- f_d)\chi_{\rm cw}(T) 
 + f_d\chi_d(T) ,
\label{chi_tot}
\end{eqnarray}
and
\begin{equation} 
    \chi_{\rm cw}(T) = \frac{C}{T-\theta_p},
\label{CW}
\end{equation}

where, $\theta_{\rm p}$ is the Weiss temperature, and $C=Ng^2s(s+1)\mu_B^2/3k_B$.

\begin{figure}
\includegraphics[width = 2.9in]{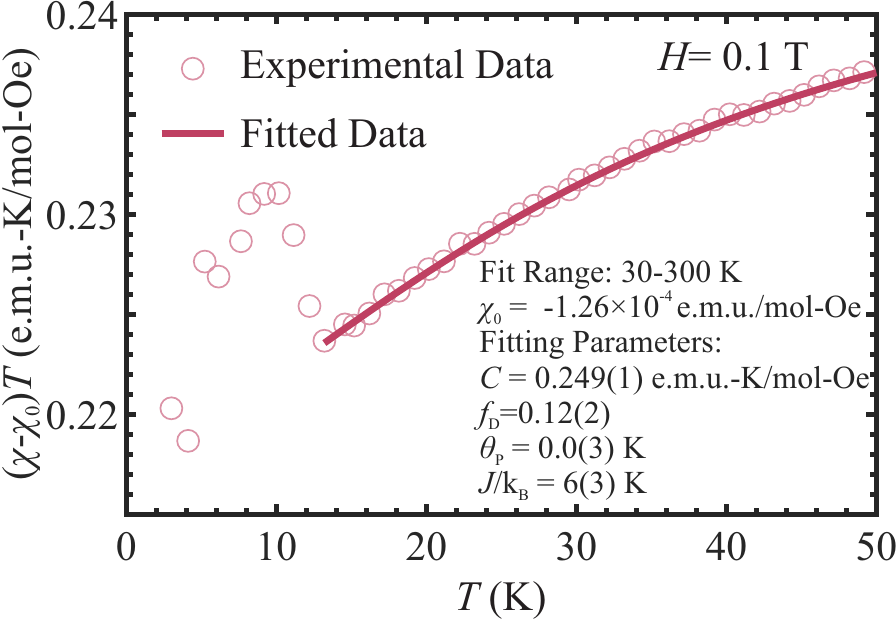}
\caption{{\bf Analysis of Susceptibility at low temperatures:} Temperature dependence of the $(\chi - \chi_0)T$ data for Sb$_{1.94}$Mn$_{0.06}$Te$_3$ along with fit following Eq.~(\ref{CW}) in the range of 13 to 50 K where the dimers prevail. $\chi_0$ is found from the fitting parameters in Figure \ref{MagSus} and the rest of the parameters listed on the figures are used as fitting parameters.}
\label{chiTvT}
\end{figure}

To account for the formation of dimers in $\chi_{\rm total}(T)$ at low temperature, we proceed by assuming a Heisenberg type interaction between two spins 
\begin{equation}
H = -J(\vec{s}_1\cdot\vec{s}_2),
\label{eqac1}
\end{equation}
where $J$ is the exchange interaction. The energy levels in terms of the total spin S=0,1,2,3,4,5 is readily given by 
\begin{eqnarray}
E(S)  = \frac{J}{2}[S(S+1) - 2s(s+1)].
\end{eqnarray}
The second term $Js(s+1)$ is constant and can be ignored .
In the presence of a magnetic field ${\bf H} = H\,\hat{\bf k}$ the energy levels are Zeeman-split into
\begin{equation}
E(S,M_S) =  \frac{J}{2}S(S+1)  + g\mu_B M_S H.
\label{eqac6}
\end{equation}
where, $M_S$ is the $z$-axis quantum number of the spin. 
To calculate the magnetic moment of the dimer in field we calculate the partition function $Z$ of the system 
\begin{equation}
Z = \sum_{S = 0}^{5}\sum_{M_S = -S}^{S} e^{-E/k_{\rm B}T},
\label{eqac7}
\end{equation}
where, $k_{\rm B}$ is Boltzman's constant.
Inserting the expression for $E$ from Eq.~(\ref{eqac6}) gives
\begin{eqnarray}
Z=\sum_{S = 0}^{5}\sum_{M_S = -S}^{S} e^{-\left[\frac{J}{2}S(S+1) + g\mu_{\rm B}M_SH\right]/k_{\rm B}T} \\ \nonumber \label{eqac8}.
\end{eqnarray}
The thermal-average moment per dimer $\mu_d$ is given by
\begin{eqnarray}
\mu_d =k_BT\frac{\partial }{\partial H}\log(Z).
\end{eqnarray}

\begin{figure}
\includegraphics[width = 2.9in]{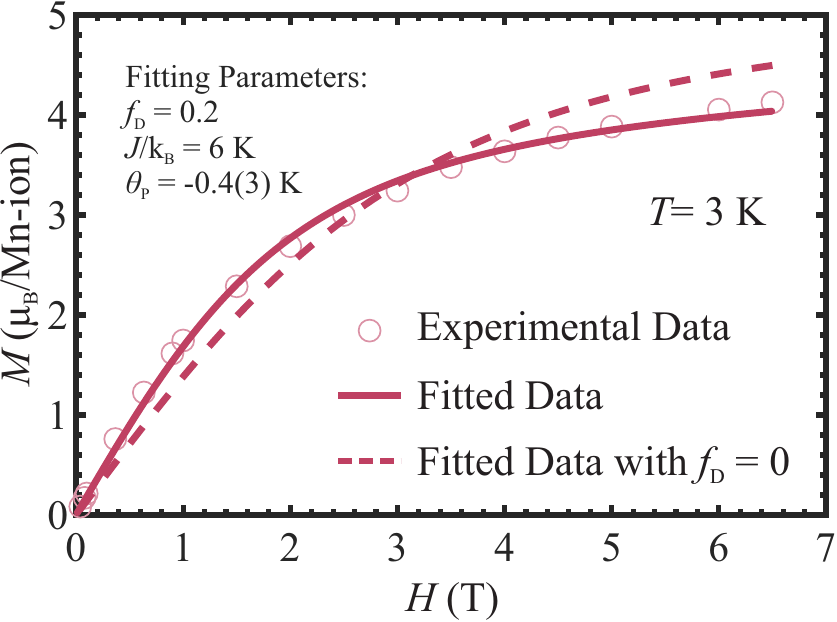}
\caption {{\bf Magnetic field dependent magnetization:} Magnetic moment per Mn in Sb$_{1.94}$Mn$_{0.06}$Te$_3$ versus applied magnetic field $H$. The solid line is best fit to the data due a combined contribution from lone and dimerized Mn atoms. An FM component on order $\sim 0.15$ $\mu_B$ is also to the fit indicating a small FM due the VdW clusters as described in the main text. The dashed line is the expected calculated magnetization  of paramagnetic Mn atoms using the Brillouin function only. }
\label{MvH}
\end{figure}

The magnetic susceptibility assuming all Mn atoms ($N$) form dimers  in the limit $g \mu_B M_S H / k_B T << 1$. With $\chi_d = \mu_d/H$, we get
\begin{equation}
\chi_d = \frac{C}{2(s(s+1) Z}\sum_{S = 0}^{5}S(S + 1)(2S + 1) e^{-\frac{J}{2}S(S+1)/k_{\rm B}T}
\label{eqac13}
\end{equation}
where $C$ is the same as that in Eq.\ \ref{CW}, assuming all $N$ Mn atoms fom dimers.

To model $\chi$ in figure \ref{MagSus}, we use Eq. (\ref{CW}) in the temperature range 30 to 300 K, for which the $f_{\rm d}$ term is negligible. This is justified, since at high temperature ($T >> J/k_B$), the dimers are coupled but uncorrelated. Indeed, our fit yields $f_{\rm d} \sim 0$. In addition, we also obtain $\chi_0=-1.26\times 10^{-4}$ emu/mole-Oe, $C=0.252(1)$ emu-K/mole-Oe, and $\theta_p=-3(4)$ K. Assuming the effective magnetic moment of Mn$^{2+}$ $\mu_{eff} = g\sqrt{s(s+1)} =  5.75$ $\mu_B$, we find that the concentration of Mn in this crystal is $p \sim 3$\%.

To emphasize the role of dimer susceptibility at low temperature, we plot $(\chi - \chi_0)T$ vs. $T$ (where $\chi_0$ is obtained from the fit above), as shown in Fig. \ref{chiTvT}. The low temperature data show an anomaly at $\sim 13$ K, which is consistent with the contribution from dimer susceptibility, as shown in Ref. \cite{Vaknin2020}. It also shows a maximum at $\sim 10$ K, which is related to FM correlation discussed in the main text. To fit the data, we use Eq. (\ref{CW}), and  find that the anomaly in the calculation due to dimers does not fit the data properly (the anomaly in the calculation is smoother). We therefore, restrict our fit to the data from 13 to 50 K. Fitting this data yields $C=0.249(2)$ e.m.u./mol-Oe, $\theta_P = 0.0(3)$ K, $f_{\rm d} = 0.12(2)$, and $J = 6(3)$ K. Note that the extracted value of $J$ is consistent with the one extracted from the INS spectra, as described in the main text.

To estimate $f_{\rm d}$, we follow the procedures developed in Ref. \cite{Behringer1958} to estimate the distribution of dimers in our crystals. Assuming $p$ is the concentration of Mn in the system and based on the symmetry of the system, we estimate that the number of dimers at the NNN inter-layer is as follows:
\begin{eqnarray}
f_d = 3p(1-p)^n,
\label{fd}
\end{eqnarray}
where $n$ is the number of excluded sites in defining an isolated dimer. For 3\% Mn ($p = 0.03$) and $n = 3$, Eq. \ref{fd} yields $f_{\rm d} \sim 0.08$, which is reasonably consistent with the analysis of susceptibility in Fig. \ref{chiTvT}.

\section{FM Spectra}

\begin{figure}[b]
\includegraphics[width = 0.9\linewidth]{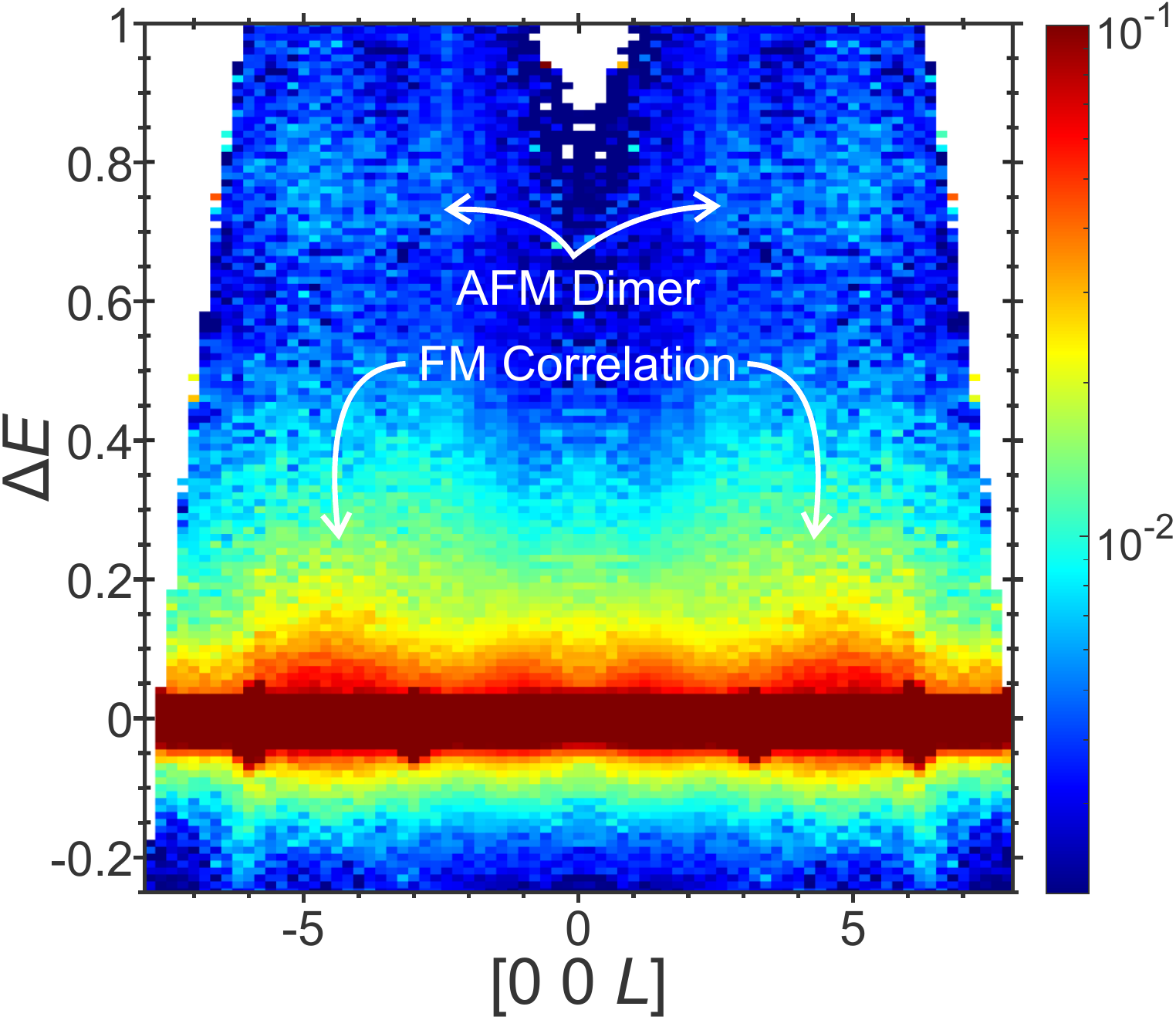}
\caption {{\bf Spectra as a function of $L$:} The spectra is obtained by integrating over $H = [-0.1, 0.1]$ and $K = [-0.1, 0.1]$ r.l.u. for $T=2$ K. The spectra concentrated at low energies ($\approx 0.1$ meV) with two humps at $L\approx \pm 5$ are due to FM correlation across the vdW gap. Possible hump at the center is truncated by the beam-stop. The couple of blobs at $\Delta E \approx 0.7$ meV originate from the AFM dimer described in the main text.}
\label{dEvsL}
\end{figure}

A different view of the spectra due to the FM correlation is shown in Fig. \ref{dEvsL}. The energy scale of the FM correlation and the FM dimers across the vdW gap can resolve specific excitation with this setup at this temperature. 

\section{DFT Results}

Table \ref{tbDFT} summarizes the calculated dimer length (DL), the magnetic moment of Mn atoms, the energy difference between AFM and FM ($\Delta E= E_\text{FM}-E_\text{AFM}$) Mn-Mn orderings, and the corresponding exchange couplings ($J$) of different Mn dimer pairs, calculated using $U=3$ eV applied on Mn-$d$ orbitals.
We find the shortest intra-layer dimer and intra-block dimer are energetically less favorable (see Fig. 4) comparing to the longer dimers.

{\it Local structural relaxation}:
Mn impurities induce a local structural distortion, although, in comparison to Bi, Sb shares a more similar atomic size with Mn.
As shown in Table \ref{tbDFT}, relaxation decreases the DL for intra-layer and intra-block dimers while increasing the DL for inter-block dimers.
This trend is consistent with experiments.
Specifically, for example, for the dominant intra-block and inter-block coupling, the unrelaxed distances are 6.40 \AA~and 9.55 \AA, respectively. 
With $U=$~{0}~--~{4}~{eV}, relaxation decreases the AF-DL by $\sim5$--$6\%$ ($\sim 6\%$ in experiments) and increase FM-DL by $\sim 2\%$ ($\sim 3\%$ in experiments).
The dependence local structural relaxation on the $U$ value is not strong, as shown in Fig. 4b.

\begin{figure}
\includegraphics[width=\linewidth]{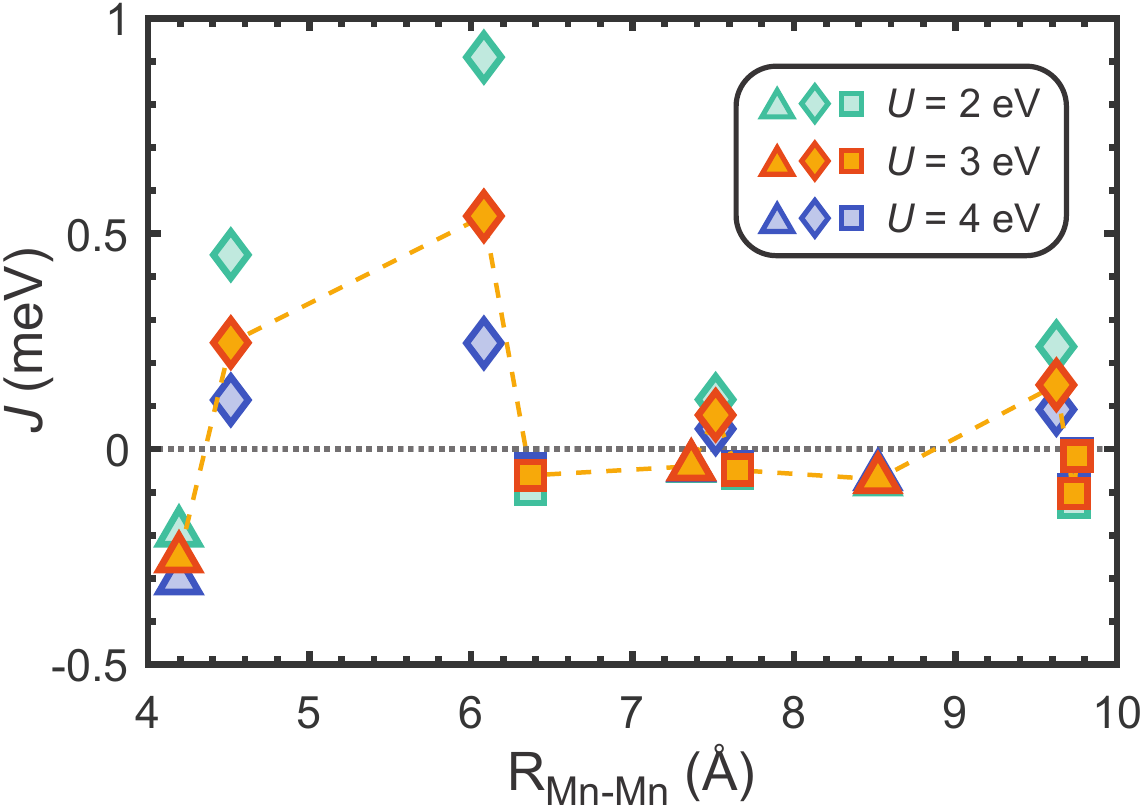}
\caption{{\bf DFT calculations:} Exchange coupling constants as a function of Mn-Mn distance calculated for $U$ = 2,3, 4 eV.}
\label{DFTSI}
\end{figure}

{\it Exchange coupling}:
Table \ref{tbDFT} shows the magnetic energies and exchange couplings calculated at $U=$~3~eV.
Interestingly, all intra-layer and inter-block couplings are FM while the intra-block couplings are AFM, which is consistent with experiments.
At $U=$~3~eV, the intra-block dimer with a {180}$^\circ$ Mn-Te-Mn bonding angle gives the strongest AFM coupling.
Table \ref{tbDFT} lists two pairs of inter-block dimers with an unrelaxed DL of {9.551}~{\AA}; one that connects through an Mn-Te-Te-Mn path and gives the strongest FM coupling, wheres the other has no obvious exchange path with intervening Te anions and gives negligible coupling.
The on-site Mn magnetic moments in all dimer configurations remain at $\sim {4.44}$~ $\mu_B$ (at $U=$~3~eV) due to the dilute nature of Mn concentration.
The Mn moment is much larger than in metallic Mn and close to the atomic limit, suggesting that the Mn is relatively localized, and the electron correlation beyond DFT may need to be considered. 

{\it $U$ dependence}:
Electron correlations can substantially affect the magnetic interactions in vdW materials~\cite{Ke2021}.
DFT+$U$ provides the simplest form to include the on-site non-local correlation effects.
However, the choice of $U$ value is ambiguous; it is worth investigating  the dependence of $J$ values on the choice of $U$ value.
Sizable $U$ values of {4}~--~{5}~{eV} in DFT+$U$ calculations give good descriptions of magnetic interactions in MnBi$_{2}$Te$_{4}$ and MnSb$_{2}$Te$_{4}$ systems~\cite{Li2020,Li2021a,Riberolles2021,Lai2021}.
One may expect that a sizable $U$ is also needed to describe the magnetic interactions in this dilute impurity system.
Figure 4(a) shows the $U$ dependence of exchange constant $J$ for dominant AFM inter-block coupling.
The relaxation makes anti-ferromagnetic interaction stronger except at $U=0$ eV.
The AFM interaction becomes weaker with the increase of $U$, and at $U=$~3~eV, the calculated $J$ is similar to the experimental value.

\begin{table}[h]
  \caption{The dimer length (DL), magnetic energy $\Delta E$, and exchange coupling $J$ values calculated for intra-layer, intra-block, and inter-block magnetic Mn-Mn dimers.
    Calculations are performed in DFT+$U$ with $U=3$ eV. Magnetic energies are calculated as $\Delta E= E_\text{FM}- E_\text{AFM}= 2J\,s^2_\text{Mn}=J\,m^2_\text{Mn}/2$. 
    Positive (negative) $\Delta E$ and $J$ values correspond to AFM (FM) couplings between the two Mn atoms of a dimer.
    The couplings between Mn pairs that have a distance larger 10 \AA~are ignored when mapping $\Delta E$ to $J$, resulting in one or two identical pairs of Mn-Mn coupling in the supercell.
  The Mn spin moment m$_{{\rm Mn}}$ = 4.44 $\mu_{\rm B}$ remains the same in all calculations.}
	\label{tbl:dis1}
	\bgroup
	\def\arraystretch{1.2}
               
        \begin{tabular*}{\linewidth}{c@{\extracolsep{\fill}}cccc}
        	\hline \hline  
        & Unrelaxed DL & Relaxed DL  & $\Delta E$ & $J$\\
        & (\AA)       & (\AA) & (meV/cell) & (meV) \\ \hline
        Intra-	      & 4.260 & 4.192      & -3.170  & -0.310 \\
        layer	      & 7.379 & 7.364      & -0.495  & -0.048 \\
        	      & 8.520 & 8.520      & -1.750  & -0.086 \footnotemark[1] \\ \hline
		      & 4.772 & 4.513      &  3.088  &  0.302 \\
	Intra-	      & 6.397 & 6.080      &  6.757  &  0.662 \footnotemark[2] \\
	block 	      & 7.686 & 7.515      &  0.993  &  0.097 \\  
		      & 9.765 & 9.625      &  3.731  &  0.183 \footnotemark[1] \\ \hline
 		      & 6.064 & 6.369      & -0.768  & -0.075 \\
 	Inter-	      & 7.411 & 7.650      & -0.608  & -0.060 \\
block\footnotemark[4] & 9.551 & 9.732      & -1.285  & -0.126 \footnotemark[3] \\  
 		      & 9.551 & 9.754      & -0.198  & -0.019 \\ \hline \hline	                
        \end{tabular*}        
	\egroup
        \footnotetext[1]{There are two identical Mn pairs within the 10 \AA~cut-off distance.}
        \footnotetext[2]{Dominant AFM intra-block coupling.}
        \footnotetext[3]{Dominant FM inter-block coupling.}
        \footnotetext[4]{For inter-block-dimer calculations, van der Waals functionals are employed. However, even without accounting for the van der Waals correction, there is no significant change in the trends that optimized structures show (See Supplementary table).}
\label{tbDFT}
\end{table}

\end{document}